\documentclass[12pt]{article}
\usepackage{psfig}
\usepackage{graphics}
\usepackage{caption2}
\begin{document}

\def\lsim{\raisebox{-.1cm}{$ \stackrel{<}{\sim}$}}
\title{\bf CP violation at a linear
collider with transverse polarization}
\date{}
\author{B. Ananthanarayan$^{1,2}$ and Saurabh D. Rindani$^3$ \\[.5cm]
\it $^1$Thomas Jefferson National Accelerator Facility \\ \it
Newport News, Virginia 23606, USA \\[.2cm] \it $^2$Centre for
High Energy Physics, Indian Institute of Science \\ \it Bangalore
560 012, India\footnote{Permanent address}\\[.2cm] \it $^3$Theory
Group, Physical Research Laboratory \\ \it Navrangpura, Ahmedabad
380 009, India} \maketitle
\begin{abstract}
We show how transverse beam polarization at $e^+e^-$ colliders can
provide a novel means to search for CP violation by observing the
distribution of a single final-state particle without measuring
its spin. We suggest an azimuthal asymmetry which singles out
interference terms between standard model contribution and
new-physics scalar or tensor effective interactions in the limit
in which the electron mass is neglected. Such terms are
inaccessible with unpolarized or longitudinally polarized beams.
The asymmetry is sensitive to CP violation when the transverse
polarizations of the electron and positron are in opposite senses.
The sensitivity of planned future linear colliders to new-physics
CP violation in $e^+e^- \to t \bar{t}$ is estimated in a
model-independent parametrization. It would be possible
to put a bound of  $\sim 7$ TeV 
on the new-physics scale $\Lambda$ at the $90\%$ C.L. for
$\sqrt{s}=500$ GeV and $\int dt {\cal L}=500\, {\rm fb}^{-1}$, with
transverse polarizations of 80\% and 60\% for the electron and
positron beams, respectively.
\end{abstract}

\newcommand{\nc}{\newcommand}
\nc{\jc}{\frac{1}{4}}  \nc{\sll}{S_{LL}}     \nc{\slr}{S_{LR}}
\nc{\srl}{S_{RL}}      \nc{\srr}{S_{RR}}     \nc{\vll}{V_{LL}}
\nc{\vlr}{V_{LR}}      \nc{\vrl}{V_{RL}}     \nc{\vrr}{V_{RR}}
\nc{\tll}{T_{LL}}      \nc{\tlrs}{T_{LR}}    \nc{\trl}{T_{RL}}
\nc{\trr}{T_{RR}}      \nc{\slld}{S_{LL}^D}  \nc{\slrd}{S_{LR}^D}
\nc{\srld}{S_{RL}^D}   \nc{\srrd}{S_{RR}^D}  \nc{\vlld}{V_{LL}^D}
\nc{\vlrd}{V_{LR}^D}   \nc{\vrld}{V_{RL}^D}  \nc{\vrrd}{V_{RR}^D}
\nc{\tlld}{T_{LL}^D}   \nc{\tlrd}{T_{LR}^D}  \nc{\trld}{T_{RL}^D}
\nc{\trrd}{T_{RR}^D}   \nc{\aqde}{\alpha_{qde}}
\nc{\alq}{\alpha_{\ell q}}        \nc{\alqp}{\alpha_{\ell q'}}
\nc{\alqt}{\alpha_{\ell q}^{(3)}} \nc{\alqtc}{\alpha_{\ell
q}^{(3)*}} \nc{\alqj}{\alpha_{\ell q}^{(1)}}
\nc{\alqjc}{\alpha_{\ell q}^{(1)*}} \nc{\aeu}{\alpha_{eu}}
\nc{\alu}{\alpha_{\ell u}} \nc{\aqe}{\alpha_{qe}}
\nc{\ber}{\begin{eqnarray*}} \nc{\enr}{\end{eqnarray*}}
\nc{\jmpb}{(1-\beta)/(1+\beta)} \nc{\wspR}{r}      \nc{\varx}{x}
\nc{\bt}{\beta}

\nc{\non}{\nonumber} \nc{\lspace}{\;\;\;\;\;\;\;\;\;\;}
\nc{\llspace}{\lspace \lspace}
\nc{\jnl}{\frac{1}{{\mit\Lambda}^2}} \nc{\jd}{\frac{1}{2}}

\nc{\ee}{e^+e^-}
\def\tt{t\bar{t}}
\nc{\bb}{\bibitem} \nc{\ra}{\rightarrow} \nc{\g}{\gamma}
\nc{\beq}{\begin{equation}} \nc{\eeq}{\end{equation}}

\def\dps{\displaystyle}

\newpage
\section{Introduction}

An $e^+e^-$ linear collider operating at a centre-of-mass (cm)
energy of a few hundred GeV and with an integrated luminosity of
several hundred inverse femtobarns is now a distinct possibility.
It is likely that the beams can be longitudinally polarized, and
there is also the possibility that spin rotators can be used to
produce transversely polarized beams. Proposals include the GLC
(Global Linear Collider) in Japan \cite{jlc}, the NLC (Next Linear
Collider) in the USA \cite{nlc}, and TESLA (TeV-Energy
Superconducting Linear Accelerator) in Germany \cite{tesla}.  The
physics objectives of these facilities include the precision study
of standard model (SM) particles, Higgs discovery and study, and
the discovery of physics beyond the standard model.

One important manifestation of new physics would be the
observation of CP violation outside the traditional setting of
meson systems, since CP violation due to SM interactions is
predicted to be unobservably small elsewhere. For instance, one
may consider the presence of model independent ``weak'' and
``electric'' dipole form factors for heavy particles such as the
$\tau$ lepton and the top quark. In case of the $\tau$ lepton, LEP
experiments have constrained their magnitudes from certain
CP-violating correlations proposed in \cite{bern}. Furthermore, it
was pointed out that longitudinal polarization of the electron
and/or positron beams dramatically improves the resolving power of
other CP-violating correlations in $\tau$-lepton \cite{ba} and
top-quark pair production \cite{cuypers}, and of decay-lepton
asymmetries in top-quark pair production \cite{pou}.

Here we consider exploring new physics via the observation of CP
violation in top-quark pair production, by exploiting the
transverse polarization (TP) of the beams at these facilities. We
rely on completely general and model-independent parametrization
of beyond the standard model interactions \cite{eichten, buch, grzad} in
terms of contact interactions, and on very general results on the
role of TP effects due to Dass and Ross \cite{dass}. We
demonstrate through explicit computations that only those
interactions that transform as tensor or (pseudo-)scalar
interactions under Lorentz transformations can contribute to
CP-violating terms in the differential cross section at
leading order when the beams have only TP. By considering
realistic energies and integrated luminosities,  and some
angular-integrated asymmetries in $e^+e^- \rightarrow t\overline
t$, we find that the scale $\Lambda$ at which new physics sets in
can be probed at the $90\%$ confidence level is $O(10)$ TeV.  This
effective scale can reach or go beyond what one might expect in
popular extensions of the SM such as the minimal supersymmetric
model, or extra-dimensional theories.  Note that the tensor and
(pseudo-)scalar interactions are accessible only at a higher order
of perturbation theory without TP, even if longitudinal
polarization is available. Also, in the foregoing, effects due to
$m_e$ are neglected everywhere.

It may be mentioned that TP in the search of new physics has
received sparse attention (for the limited old and recent
references with or without CP violation, see \cite{rizzo}). In the
CP violation context, the only work of relevance, to our knowledge,
is that of Burgess and Robinson \cite{burgess}, who considered
pair production of leptons and light quarks in the context of LEP
and SLC. Our discussion of top pair production, which is in the
context of much higher energies, does have some features in common
with the work of ref. \cite{burgess}, though the numerical
analysis is necessarily different. Furthermore, we have included a
discussion of CP violation for a general inclusive process.

In the process $e^+e^- \rightarrow f\overline{f}$, where $\overline{f}$ is 
different from $f$, testing  CP
violation needs more than just the momenta of the particles to be
measured. In the CM frame, there are only two vectors,
$\vec{p}_{e^-}-\vec{p}_{e^+}$ and $\vec{p}_{f}-\vec{p}_{\bar{f}}$.
The only scalar observable one can construct out of these is $
(\vec{p}_{e^-} - \vec{p}_{e^+}) \cdot (
\vec{p}_{f}-\vec{p}_{\bar{f}})$. This is even under CP. Hence one
needs either initial spin or final spin to be observed. Observing
the final spin in the case of the top quark is feasible because of
the fact that the top quark decays before it hadronizes. Several
studies have been undertaken to make predictions for the
polarization, and for the distributions of the decay distributions
in the presence of  CP  violation in top production and decay.

On the other hand, the presence of TP of the beams would provide
one more vector, making it possible to observe  CP  violating
asymmetries without the need to observe final-state polarization.
This would mean gain in statistics. Thus, possible  CP-odd scalars
which can be constructed out of the available momenta and TP are $
(\vec{p}_{e^-}-\vec{p}_{e^+})\times (\vec{s}_{e^-}-\vec{s}_{e^+})
\cdot ( \vec{p}_{f}-\vec{p}_{\bar{f}})$ and
$(\vec{s}_{e^-}-\vec{s}_{e^+}) \cdot (
\vec{p}_{f}-\vec{p}_{\bar{f}})$, together with combinations of the
above with CP-even scalar products of vectors. 

It may be noted that if $\overline f = f$, that
is, if $f$ is a self-conjugate boson, or a Majorana fermion,
possible CP-odd scalars in a process $e^+e^- \rightarrow f + X$
are $(\vec{p}_{e^-}-\vec{p}_{e^+})\cdot \vec{p}_{f}$, 
$(\vec{s}_{e^-}+\vec{s}_{e^+}) \cdot \vec{p}_{f}$
and
$ (\vec{p}_{e^-}-\vec{p}_{e^+})\times
(\vec{s}_{e^-}+\vec{s}_{e^+}) \cdot \vec{p}_{f}$. 
Of these, observation of the first does not need initial-state polarization,
observation of the 
second is possible with either longitudinally or transversely polarized beams, 
and the third requires $e^+$ and $e^-$ transverse polarizations.

We investigate below how new physics could give rise to such
CP-odd observables in the presence of TP of the beams, and how the
sensitivity of such measurements would compare with the
sensitivity to other observables involving TP, or final-state
polarization. While our general considerations are valid for any
one-particle inclusive final state $A$ in $e^+e^- \rightarrow
A+X$, for a concrete illustration we consider the specific process
$e^+e^- \rightarrow t\overline t$.

One may gain an insight from the elegant and general results of
Dass and Ross \cite{dass}, who listed all possible single-particle
distributions from the interference of the electromagnetic
contribution with S (scalar), P (pseudo-scalar), T (tensor), V
(vector) and  A (axial-vector) type of neutral current
interactions in the presence of arbitrary beam polarization. It
may be concluded from the tables in \cite{dass} that with only TP,
V and A coupling at the $e^+e^-$ vertex cannot give rise to
CP-violating asymmetries.  Even on generalization to include
interference of the $Z$ contribution, we have checked that the
same negative result holds. This is true so long as the $e^+e^-$
couple to a vector or axial vector current, even though the
coupling of the final state is more general, as for example, of
the dipole type. However, S, P and T can give  CP-odd
contributions like the ones mentioned earlier. These results may
also be deduced from some general results for azimuthal
distributions give by Hikasa \cite{hikasa}

For vanishing electron mass, S, P, and T couplings at the $e^+e^-$
vertex are helicity violating, whereas V and A couplings are
helicity conserving. So with arbitrary longitudinal polarizations,
they do not give any interference. Hence new physics appears only
in terms quadratic in the new coupling. However, with TP, these
interference terms are non-vanishing, and can be studied. Thus, TP
has the distinct advantage that it would be able to probe
first-order contributions to new physics appearing as S, P and T
couplings, in contrast with the case of no polarization, or
longitudinal polarization, which can probe only second order
contribution from new physics.

\section{The process $e^+e^-\to t\bar{t}$}

We now consider the specific process $e^+e^-\to t\bar{t}$. For our
purposes, we have found it economical to employ the discussion 
of ref. \cite{grzad}, based on the notation and 
formalism of \cite{buch}.
The following operators contribute
to this process \cite{grzad}:
\begin{equation}\label{operators}
\begin{array}{lcl}
{\cal O}^{(1)}_{\ell q}&\!\!=&\!\!\dps\frac12
(\bar{\ell}\gamma_{\mu}\ell)(\bar{q}\gamma^{\mu}q),\\ && \\ {\cal
O}^{(3)}_{\ell q}&\!\!=&\!\!\dps\frac12
(\bar{\ell}\gamma_{\mu}\tau^I\ell) (\bar{q}\gamma^{\mu}\tau^Iq),\\
\vspace*{-0.3cm} & &  \\ \vspace*{-0.3cm} {\cal O}_{eu}&\!\!
=&\!\!\dps\frac12 (\bar{e}\gamma_{\mu}e)(\bar{u}\gamma_{\mu}u),
\\ & &  \\ {\cal O}_{\ell u}&\!\!=&\!\!(\bar{\ell}u)(\bar{u}\ell),
\\ {\cal O}_{qe}&\!\!=&\!\!(\bar{q}e)(\bar{e}q),\\ {\cal O}_{\ell
q}&\!\!=&\!\!(\bar{\ell}e)\epsilon (\bar{q}u), \\ {\cal O}_{\ell
q'}&\!\!=&\!\!(\bar{\ell}u)\epsilon (\bar{q}e),
\end{array}
\end{equation}
where $l,q$ denote respectively the left-handed electroweak $SU(2)$
lepton and quark doublets, and $e$ and $u$ denote $SU(2)$ singlet
charged-lepton and up-quark right-handed fields. $\tau^I$ $(I=1,2,3)$ are
the usual Pauli matrices, and $\epsilon$ is the $2\times 2$ 
anti-symmetric matrix, $\epsilon_{12}=-\epsilon_{21}=1$. Generation indices
are suppressed.
The Lagrangian which we use in our
following calculations is written in terms of the above operators as 
\cite{grzad}:
\begin{equation}\label{lag}
{\cal L}={\cal L}^{S\!M}+
\frac{1}{{\mit\Lambda}^2}\sum_i(\:\alpha_i{\cal O}_i+{\rm
h.c.}\:),
\end{equation}
where $\alpha$'s are the coefficients which parameterize
non-standard interactions. Such an effective interaction could
arise in extensions of SM like multi-Higgs doublet models,
supersymmetric standard model through loops involving heavy
particles or theories with large extra dimensions.

After Fierz transformation the Lagrangian containing the
new-physics four-Fermi operators takes the form
\begin{equation}\label{lag4f}
{\cal L}^{4F}
 =\sum_{i,j=L,R}\Bigl[\:S_{ij}(\bar{e}P_ie)(\bar{t}P_jt)
 +V_{ij}(\bar{e}\gamma_{\mu}P_ie)(\bar{t}\gamma^{\mu}P_jt)
 +T_{ij}
 (\bar{e}\frac{\sigma_{\mu\nu}}{\sqrt{2}}P_ie)
(\bar{t}\frac{\sigma^{\mu\nu}}{\sqrt{2}}P_jt)\:\Bigr]
\end{equation}
with the coefficients satisfying:
\begin{eqnarray*}
&&S_{RR}=S^{*}_{LL},\ \ \ S_{LR}=S_{RL}=0,\\ &&\ \ \ \ \ \ \ \ \ \
\ \ V_{ij}=V^{*}_{ij},\\ &&T_{RR}=T^{*}_{LL},\ \ \
T_{LR}=T_{RL}=0.
\end{eqnarray*}
In (\ref{lag4f}), $P_{L,R}$ are respectively the left- and right-chirality
projection matrices.
The relation between the coefficients in eq. (\ref{lag4f}) and the
coefficients $\alpha_i$ of eq. (\ref{lag}) may be found in
\cite{grzad}. In the above scalar as well as pseudo-scalar
interactions are included in a definite combination. Henceforth,
we will simply use the term scalar to refer to this combination
of scalar and pseudoscalar couplings. 

As mentioned earlier, interference between scalar-tensor and
SM interactions can only arise in the presence of TP.
SM amplitude can, of course, interfere with
contributions from vector four-Fermi operators, leading to terms
of order $\alpha_i s/{\mit\Lambda}^2$. However, as far as CP
violation is concerned, this interference between SM amplitude and
vector amplitude from new physics does not give  CP-odd terms in
the distribution, even when TP is present. On the other hand, in
the presence of TP, the interference between SM amplitude and
scalar or tensor contribution does produce CP-odd variables.

Here we concentrate on the process $e^+e^-\rightarrow t \overline
t$ and examine the CP-violating contribution in the interference
of the SM amplitude with the scalar and tensor four-Fermi
amplitudes. We will take the electron TP to be 100\% and along the
positive or negative $x$ axis, and the positron polarization to be
100\%, parallel or anti-parallel to the electron polarization.
The $z$ axis is chosen along the direction of the $e^-$. The
differential cross sections for $e^+e^-\rightarrow t \overline t$,
with the superscripts denoting the respective signs of the $e^-$
and $e^+$ TP,  are
\begin{equation}\label{diffcspp}
\displaystyle{\frac{d\sigma^{\pm\pm}}{d\Omega}}=\displaystyle{
\frac{d\sigma^{\pm\pm}_{SM}}{d\Omega}} \mp\displaystyle{
\frac{3\alpha\beta^2}{4\pi}\!\frac{m_t\sqrt{s}}{
s-m_Z^2}}\left(c_V^tc_A^e {\rm Re} S\right) \sin\theta\cos\phi,
\end{equation}
\begin{equation}\label{diffcspm}
\displaystyle{\frac{ d\sigma^{\pm\mp}}{d\Omega}}=\displaystyle{
\frac{d\sigma^{\pm\mp}_{SM}}{d\Omega}} \pm \displaystyle{
\frac{3\alpha\beta^2}{4\pi}\!\frac{m_t\sqrt{s}}{
s-m_Z^2}}\left(c_V^tc_A^e
 {\rm Im} S\right) \sin\theta\sin\phi,
\end{equation}
where
\begin{eqnarray}\label{diffcsSM}
\frac{d\sigma^{+\pm}_{SM}}{d\Omega}&=&\frac{d\sigma^{-\mp}_{SM}}{d\Omega}\non\\
 &=&\frac{3\alpha^2\beta}{4s}
\left[ \frac{4}{9}\left\{ 1+\cos^2\theta +
\frac{4m_t^2}{s}\sin^2\theta \pm \beta^2 \sin^2\theta \cos 2\phi
\right\} \right. \non \\ &-&\!\!\!\!\!\! \left.
\frac{s}{s-m_Z^2}\,\frac{4}{3} \left\{\!c_V^e c_V^t
(1+\cos^2\theta \!+ \! \frac{4m_t^2}{s}\sin^2\theta \pm \beta^2
\sin^2\theta \cos 2\phi)\! \right. \right. \non \\ &+&\!\!\!\!
\left.\left. 2\ c_A^ec_A^t\beta\cos\theta \right\} +
\frac{s^2}{(s-m_Z^2)^2} \left\{ (c_V^{e\ 2} + c_A^{e\ 2}) \right.
\right. \non \\ &\times &\!\!\!\! \left. \left.\!\!\! \left[
(c_V^{t\ 2} + c_A^{t\ 2}) \beta^2 (1+\cos^2\theta ) +  c_V^{t\
2}\frac{8m_t^2}{s}\right]\! +8c_V^ec_A^ec_V^tc_A^t \beta
\cos\theta
 \right. \right.
\non \\ &\pm&\!\! \!\!\left. \left.\!\!\!
 (c_V^{e\ 2}-c_A^{e\ 2})
(c_V^{t\ 2} + c_A^{t\ 2}) \beta^2 \sin^2\theta\! \cos
2\phi\right\}\right]
\end{eqnarray}
Here $\beta = \sqrt{1-4m_t^2/s}$, and we have defined
\begin{equation}\label{S}
S\equiv S_{RR} + \frac{2c_A^t c_V^e}{ c_V^t c_A^e} T_{RR},
\end{equation}
where $c_V^i$, $c_A^i$ are the couplings of $Z$ to $e^-e^+$ and $t
\overline t$, and where we have retained the new couplings to
linear order only. In (\ref{S}) the contribution of the tensor
term relative to the scalar term is suppressed by a factor
$2c_A^tc_V^e/c_V^tc_A^e \approx 0.36$. In what follows, we will
consider only the combination $S$, and not $S_{RR}$ and $T_{RR}$
separately.

The differential cross section corresponding to anti-parallel
$e^-$ and $e^+$ polarizations, eq. (\ref{diffcspm}), has the
CP-odd quantity
\[
\sin\theta\sin\phi \equiv \dps \frac{(\vec p_{e^-} - \vec
p_{e^+})\times (\vec s_{e^-} - \vec s_{e^+})\cdot (\vec p_{t} -
\vec p_{\bar t}) }{\vert \vec p_{e^-} - \vec p_{e^+}\vert \vert
\vec s_{e^-} - \vec s_{e^+} \vert \vert \vec p_{t} - \vec p_{\bar
t}\vert  },
\]
while the interference term in the case with parallel $e^-$ and
$e^+$ polarizations, eq. (\ref{diffcspp}), has the CP-even
quantity
\[
\sin\theta\cos\phi \equiv \dps \frac{(\vec p_{t} - \vec p_{\bar
t})\cdot (\vec s_{e^-} + \vec s_{e^+})}{2\vert \vec p_{t} - \vec
p_{\bar t}\vert }
\]

We construct the CP-odd asymmetry, which we call the up-down
asymmetry as

\beq\label{asym} A(\theta)= \dps{ \frac{ \dps{\int_0^\pi \frac{
d\sigma^{+-}}{d\Omega}} d\phi
 - \int_{\pi}^{2\pi} \dps{\frac{ d\sigma^{+-}}{d\Omega}} d\phi
}{
 \dps{\int_{0}^{\pi} \frac{ d\sigma^{+-}}{d\Omega}} d\phi
 + \int_{\pi}^{2\pi}\dps{ \frac{ d\sigma^{+-}}{d\Omega}} d\phi
}  } \eeq and also the $\theta$-integrated version,
\beq\label{asymcutoff} A(\theta_0)= \dps{ \frac{
\dps{\int_{-\cos\theta_0}^{\cos\theta_0} \int_0^\pi \frac{
d\sigma^{+-}}{d\Omega}} d\cos\theta d\phi
 - \int_{-\cos\theta_0}^{\cos\theta_0}\int_{\pi}^{2\pi} \dps{\frac{
d\sigma^{+-}}{d\Omega}}d\cos\theta d\phi }{
 \dps{\int_{-\cos\theta_0}^{\cos\theta_0}\int_{0}^{\pi} \frac{
d\sigma^{+-}}{d\Omega}}d\cos\theta d\phi
 + \int_{-\cos\theta_0}^{\cos\theta_0}\int_{\pi}^{2\pi}\dps{ \frac{
d\sigma^{+-}}{d\Omega}}d\cos\theta d\phi }  } \eeq In the latter,
a cut-off on $\theta$ has been introduced, so that the limits of
integration for $\theta$ are $\theta_0 < \theta < \pi - \theta_0$.
Using our expressions for the differential cross sections, it is
easy to obtain expressions for these asymmetries, and we do not
present them here. Such a cut-off in the forward and backward
directions is indeed needed for practical reasons to be away from
the beam pipe. We can further choose the cut-off to optimize the
sensitivity of the measurement.

\section{Numerical results}

We now proceed with a numerical study of these asymmetries and the
limits that can be put on the parameters using the integrated
asymmetry $A(\theta_0)$. We assume that a linear collider
operating at $\sqrt{s}=500$ GeV and the ideal condition of 100\%
beam polarizations for $e^-$ as well as $e^+$. We will comment
later on about the result for more realistic polarizations.
\begin{figure}[!hb]
\centering \psfig{file=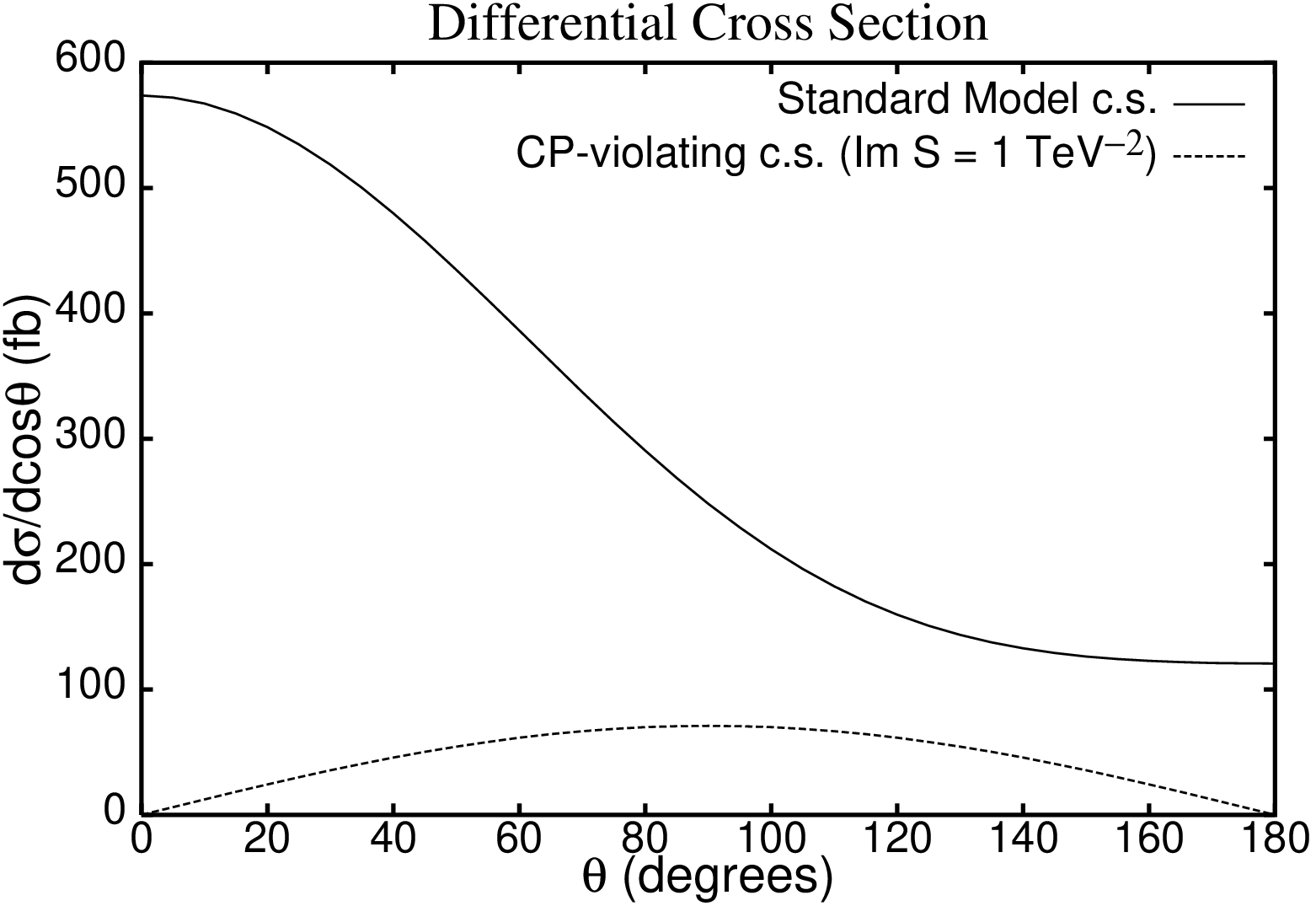,width=4.2in,angle=-90}
\caption{The SM differential cross section
$\frac{d\sigma}{d\cos\theta}$ (fb) (solid line) and the numerator
of the asymmetry $A(\theta)$ in eq. (\ref{asym}) (broken line) as
a function of $\theta$. The latter is for Im $S=1$ TeV$^{-2}$. }
\centering \psfig{file=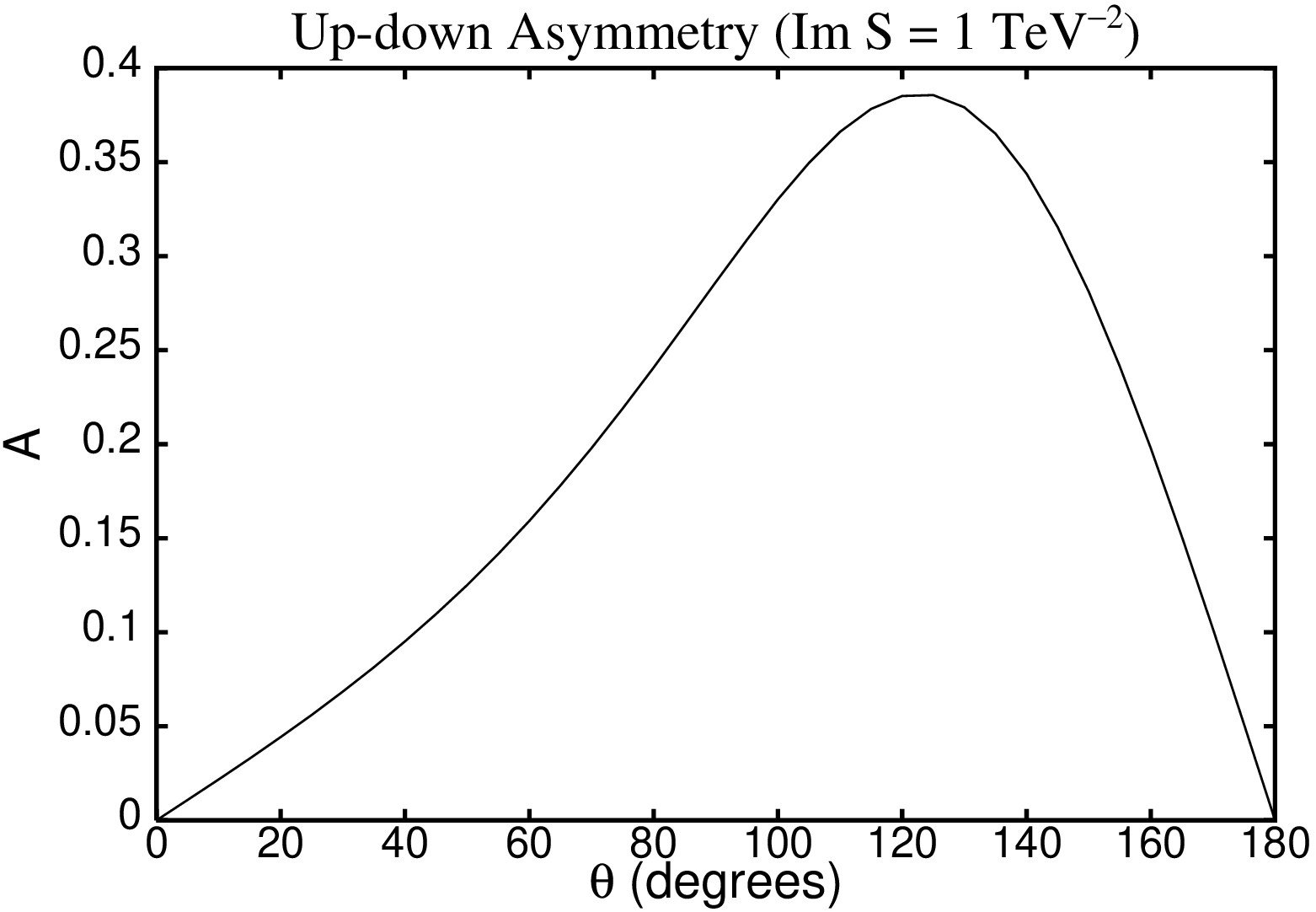,width=4.2in,angle=-90} \caption{The
asymmetry $A(\theta)$ defined in eq. (\ref{asym}) as a function of
$\theta$ for a value of Im $S=1$ TeV$^{-2}$.}
\end{figure}

\begin{figure}[!htb]
\centering \psfig{file=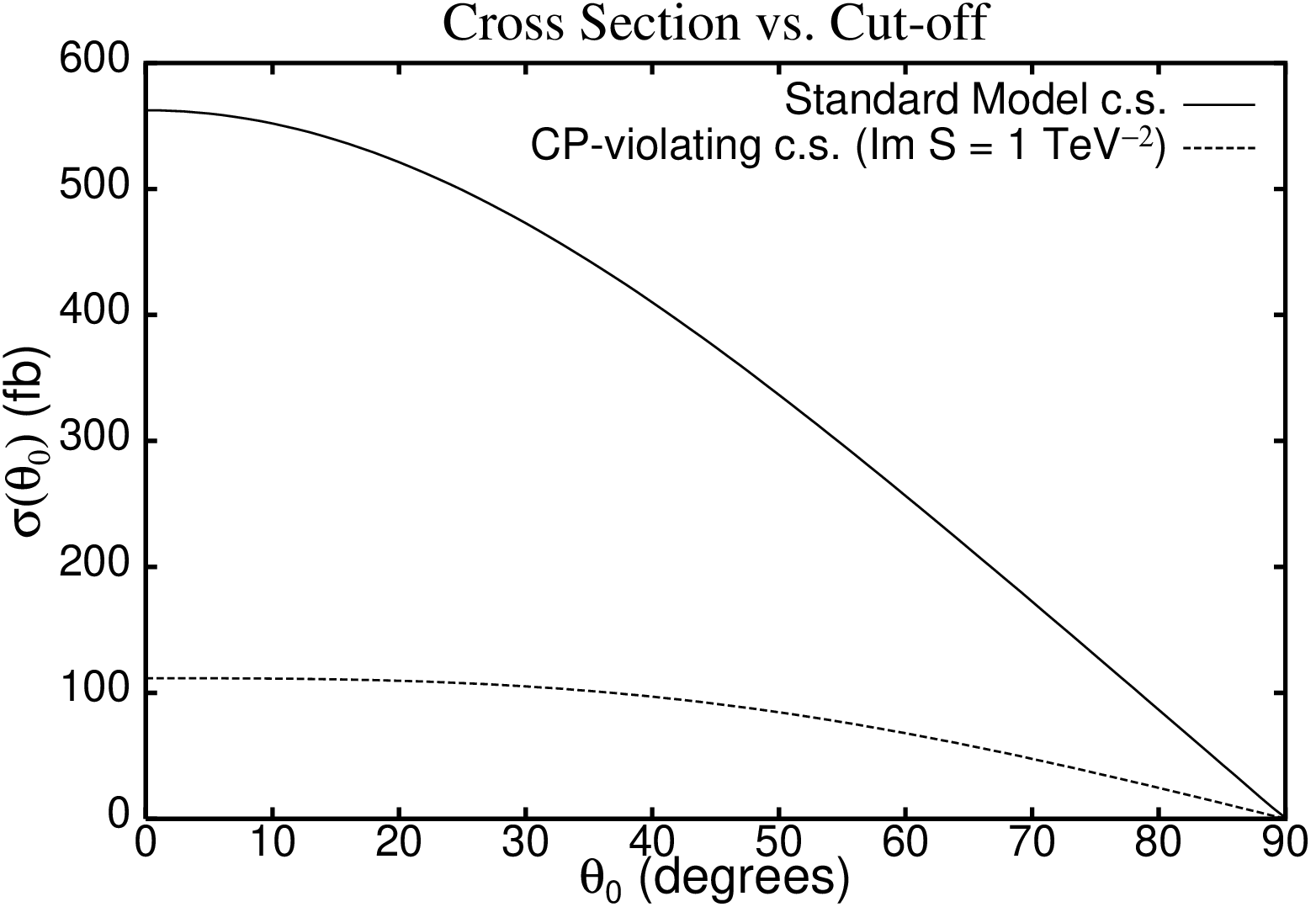,width=4.3in,angle=-90}
\caption{As in Fig. 1, but now for quantities integrated over
$\theta$ with a cutoff $\theta_0$,  plotted as a function of
$\theta_0$.}
\centering \psfig{file=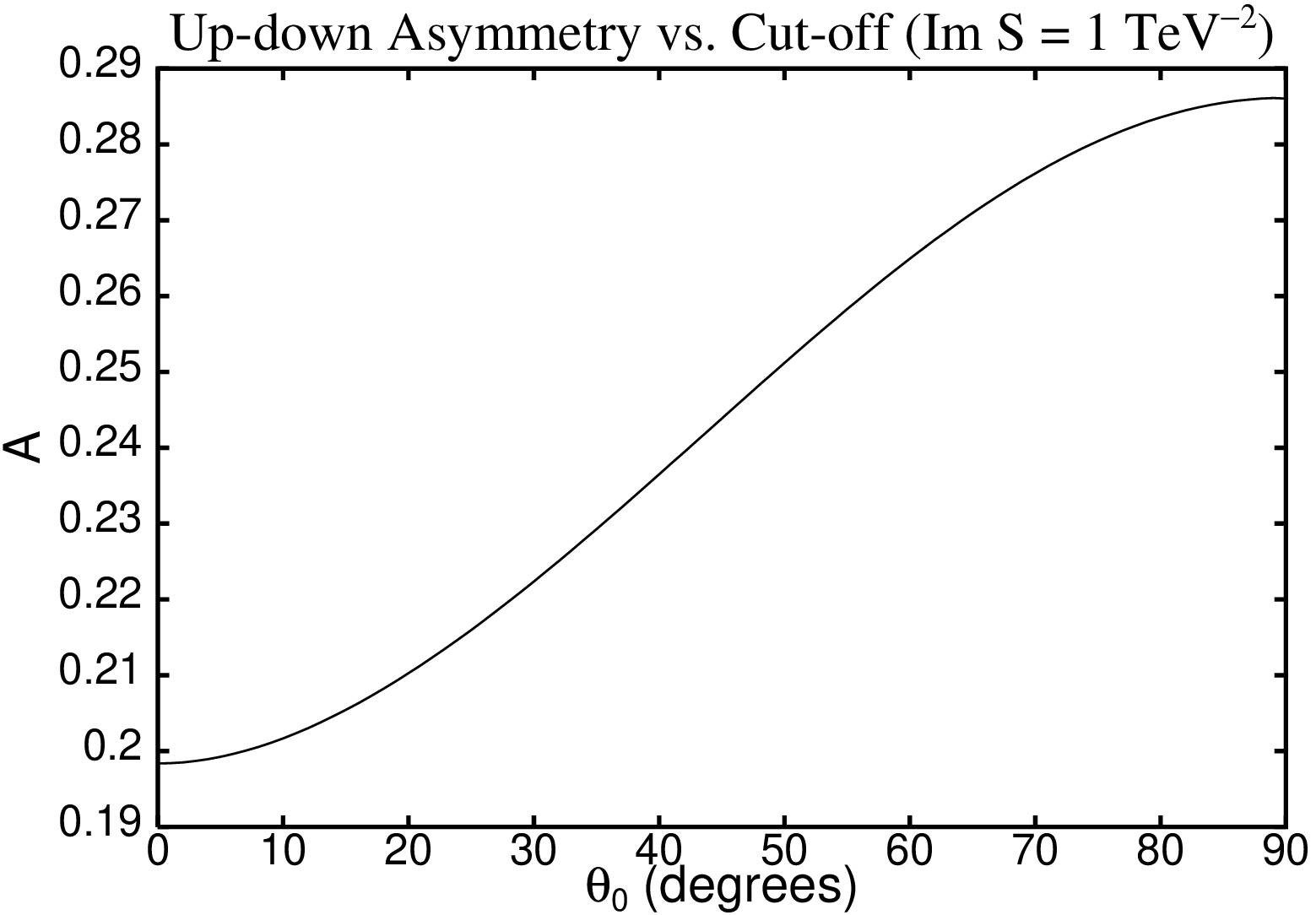,width=4.3in,angle=-90}
\caption{The asymmetry $A(\theta_0)$ defined in eq.
(\ref{asymcutoff}) plotted as a function of $\theta_0$ for Im
$S=1$ TeV$^{-2}$.}
\end{figure}

In Fig. 1, we plot the SM differential cross section integrated
over $\phi$, as well as the numerator of $A(\theta)$ of eq.
(\ref{asym}), which is two times the contribution of the
interference term (the CP-violating contribution coming from ${\rm
Im}S$) integrated over $\phi$ from 0 to $\pi$ for a value of ${\rm
Im}S= 1$ TeV$^{-2}$. In Fig. 2 we show the asymmetry $A(\theta)$
for ${\rm Im}S= 1$ TeV$^{-2}$ as a function of $\theta$. The
asymmetry peaks at about $\theta = 120^\circ$, and takes values as
high as 30-40\%.

\begin{figure}[htb]
\begin{center}
\centering
\centerline{\psfig{file=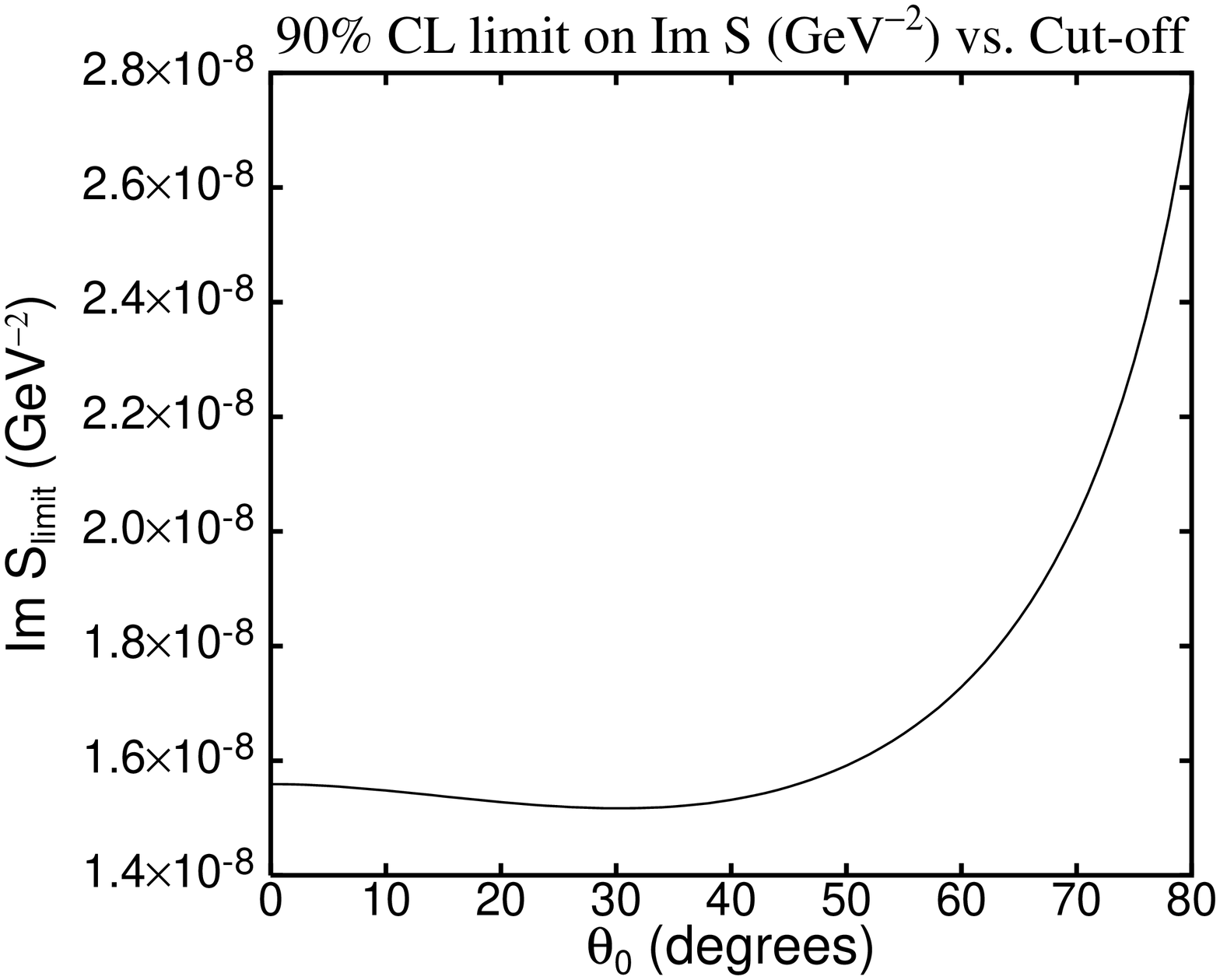,width=4in,angle=-90}}
\caption{The 90\% C.L. limit that can be obtained on Im $S$ with
an integrated luminosity of 500 fb$^{-1}$ plotted as a  function
of the cut-off angle $\theta_0$.}
\end{center}
\end{figure}

In Fig. 3 we plot, as functions of the cut-off angle $\theta_0$, 
the $\theta$-integrated versions of the quantities that
are plotted in Fig. 1.  The limits of integration are
$\theta_0$ and $\pi - \theta_0$.
Fig. 4 shows the integrated
up-down asymmetry $A(\theta_0)$ as a function of $\theta_0$. The
value of $A(\theta_0)$ increases with the cut-off, because the SM
cross section in the denominator of eq. (\ref{asymcutoff})
decreases with cut-off faster than the numerator.

Fig. 5 shows the 90\% confidence level (C.L.) limits that could
be placed on ${\rm Im}S$ for an integrated luminosity of $L=500$
fb$^{-1}$. The limit is the value of Im $S$ which would give rise
to an asymmetry $A_{\rm lim} =1.64/\sqrt{L\Delta\sigma}$, where
$\Delta\sigma$ is the SM cross section. The limit is relatively
insensitive to the cut-off $\theta_0$ until about
$\theta_0=60^{\circ}$, after which it increases. A cut-off could
be chosen anywhere upto this value. The corresponding limit is
about $1.6\cdot 10^{-8}$ GeV$^{-2}$, after which it gets worse.
This limit translates to a value of $\Lambda$ of the order of 8
TeV, assuming that the coefficients $\alpha_i$ in (\ref{lag}) are
of order 1. The corresponding limit for $\sqrt{s}$ of 800 GeV with the
same integrated luminosity is $\sim 9.5$ TeV.

So far we have assumed 100\% TP for both $e^+$ and $e^-$ beams. We
now discuss the effect of realistic TP. Since longitudinal
polarizations of 80\% and 60\% are likely to be feasible
respectively for $e^-$ and $e^+$ beams, we will assume that the
same degree of TP will also be possible. 
We are assuming here that spin rotators that convert longitudinal polarization
 to TP will not deplete
the degree of polarization significantly.
Since we use differential
cross sections integrated over $\phi$ at least over the range 0 to
$\pi$, the polarization dependent terms in the SM contribution,
being proportional to $\cos2\phi$, drop out, as does the second
term on the right-hand side of eq. (\ref{diffcspp}). So far as the
up-down asymmetry $A(\theta)$ or $A(\theta_0)$ is concerned, it
gets multiplied by a factor $\frac{1}{2}(P_1 - P_2)$ in the
presence of degrees of TP $P_1$ and $P_2$ for $e^-$ and $e^+$
beams respectively. For $P_1=0.8$ and $P_2=-0.6$, this means a
reduction of the asymmetry by a factor of 0.7. Since the SM cross
section does not change, this also means that the limit on the
parameter Im $S$ goes up by a factor of $1/0.7\approx 1.4$, and
the limit on $\Lambda$ goes down by a factor of $\sqrt{0.7}\approx
0.84$, to about 6.7 TeV. If the positron beam is unpolarized,
however, the sensitivity goes down further.


\section{Conclusions}

In summary, TP can be used to study CP-violating asymmetry arising
from the interference of new-physics scalar and tensor
interactions with the SM interactions. These interference terms
cannot be seen with longitudinally polarized or unpolarized beams.
Moreover, such an asymmetry would not be sensitive to new vector
and axial-vector interactions (as for example, from an extra $Z'$
neutral boson), or even electric or ``weak" dipole interactions of
heavy particles, since the asymmetry vanishes in such a case in
the limit of vanishing electron mass. Since the asymmetry we
consider does not involve the polarization of final-state
particles, one expects better statistics as compared to the case
when measurement of final-state polarization is necessary.

We have studied the  CP-violating up-down asymmetry in the case of
$e^+e^- \rightarrow t \overline t$ in detail using a
model-independent parametrization of new interactions in terms of
a four-Fermi effective Lagrangian. We find that a linear collider
operating at $\sqrt{s}=500$ GeV with an integrated luminosity of
500 fb$^{-1}$ would be sensitive to CP-violating new physics scale
of about 8 TeV corresponding to a four-Fermi coupling of about
$1.6\cdot 10^{-8}$ GeV$^{-2}$ with fully polarized beams, and
somewhat lower scales if the polarization is not 100\%.

Present experimental limits on the scale of 
CP-conserving new physics interactions are of the same order or better
than those
obtainable for CP-violating interactions as described above.
Limits of order 10-20 TeV have been obtained 
for production of light quarks \cite{cheung}.
The limits are somewhat lower for
rare flavour-violating processes \cite{black}. Recently Rizzo
\cite{rizzo2} has discussed the dependence on linear collider
energies and luminosities and on positron polarization of the
reach of future experiments on contact interaction searches. Our
discussion, while not as exhaustive, extends this in another
direction, namely, that of CP violation in the presence of TP.

While it is clear that scalar and tensor effective four-fermion
interactions can arise in many extensions of the standard model,
definite predictions of their magnitudes are, to our knowledge,
not available. However, it is likely that CP-violating box
diagrams, which seem to contribute significantly in supersymmetric
theory (see for example \cite{guasch}), may lead to such effective
interactions in many extensions of SM. One obvious case where a
tensor contribution occurs is when one includes a CP-violating
dipole coupling of the electron to $\gamma$ and $Z$, and one does
expect azimuthal asymmetries in the presence of transverse
polarization \cite{hoo}. However, in view of the strong limits on
the electric dipole moment of the electron, the effect will be
tiny, and we have not considered it here.

One may ask if there are any naturalness constraints on the parameters of the
Lagrangian ${\cal L}$. 
It is possible to conclude that for the effective theory a
constraint may arise from requiring that the one-loop contribution $\delta m_e$
to the electron mass due to scalar interactions is small compared to $m_e$.
The electron mass shift would be proportional to the square of the cutoff:
\begin{equation}
\delta m_e \sim \frac{m_t}{8\pi^2}{\rm Re}~S_{RR}\Lambda^2\;.
\end{equation}
It should be noted that such a contribution, if the underlying theory is
renormalizable, would be renormalized to zero, and so would not arise in the
underlying theory. Secondly, this contribution is independent of Im $S_{RR}$,
which is sought to be constrained from CP-odd asymmetry. 
 However, if from the point of view of naturalness one
requires $\delta m_e < m_e$, then the conclusion would be that
 the ${\rm Re} S_{RR}$ must be
suppressed by an additional factor $8\pi^2 m_e/m_t \sim 10^{-4}$,
independent of the new physics scale $\Lambda$.

Constraints could arise on the
magnitude of the tensor coupling due to possible contributions to the electron
electric and magnetic dipole moments. For example, the real part of the tensor
coupling would contribute an amount
\begin{equation}
\frac{m_e m_t}{16\pi^2}{\rm Re}~T_{RR}\; \log{(\Lambda^2/m_t^2)}
\end{equation}
to the $g-2$ of the electron, which would give a constraint 
$${\rm Re}~T_{ij} \lsim
10^{-3}~{\rm TeV}^{-2}$$
when we impose the requirement that the additional contribution is less than
the experimental uncertainty of about $8\times 10^{-12}$.
The imaginary part of the tensor coupling would contribute an amount
\begin{equation}
\frac{m_t}{16\pi^2}{\rm Im}~T_{RR}\; \log{(\Lambda^2/m_t^2)}
\end{equation}
to the electric dipole moment $d_e$ of the electron, giving a constraint 
$${\rm Im}~T_{ij}\lsim
10^{-8}~{\rm TeV}^{-2}$$
when we impose the experimental constraint of $d_e\lsim 10^{-27}~e$ cm.
However, there is always the possibility
of cancellations between contributions from different four-Fermion couplings,
only one of which (viz., the one corresponding to the $t\bar t$ final state) 
contributes to $\ee \to \tt$. While such a cancellation may seem ``unnatural",
we take the point of view that constraints on individual couplings can only be
obtained from direct experimental study of processes like $\ee \to \tt$,
including the proposal discussed in this work. 
Note, however, that from all the above, Im $S_{RR}$ remains completely
unconstrained.

We have restricted ourselves mainly to the $\sqrt{s}$ value of 500 GeV. A
linear collider operating at other energies would give similar
results. In terms of the new physics scale $\Lambda$, it is
expected that colliders at higher energies would be able to put a
better limit on $\Lambda$, since the new interactions would be
enhanced relative to SM for larger $\sqrt{s}$, indeed as we have illustrated
for the case of $\sqrt{s}=800$ GeV. In this work, we
have combined many simple principles which in our opinion make the
results of this investigation particularly compelling.

\medskip

This work is supported by the Department of Energy under contract
DE-AC-05-84ER40150 and the Department of Science and Technology
(DST), Government of India. BA thanks J. Gasser for discussions. 
SDR thanks the DST for support under
project number SP/S2/K-01/2000-II, A. Joshipura and S.Y. Choi
for discussions,
and A. Bartl and E. Christova for useful correspondence.

\end{document}